# High Security Image Steganography with Modified Arnold's Cat Map

Minati Mishra
Fakir Mohan University
Balasore,India
minatiminu@yahoo.com

Ashanta Ranjan Routray
Fakir Mohan University
Balasore,India
ashan2r@yahoo.co.in

Sunit Kumar
J. C. College,
Kolhan University, Jharkhand
sunit.dba@gmail.com

## ABSTRACT
Information security is concerned with maintaining the secrecy, reliability and accessibility of data. The main objective of information security is to protect information and information system from unauthorized access, revelation, disruption, alteration, annihilation and use. This paper uses spatial domain LSB substitution method for information embedding and Arnold's transform is applied twice in two different phases to ensure security. The system is tested and validated against a series of standard images and the results show that the method is highly secure and provides high data hiding capacity.

## 1. INTRODUCTION

Steganography is an art of hiding information in ways that prevent the detection of hidden messages and this is achieved by hiding a piece of information inside another piece of innocent looking information. There exist a number of data embedding methods such as the spatial and time domain methods, Transform domain methods and fractal encoding methods etc. These methods hide/embed information in different types of media such as text, image, audio, video etc. Amongst these varieties of different file formats, digital images are considered to be the most popular type of carriers because of their size and distribution frequency. Covert or hidden communication is the process of hiding a piece of information in another information [1]. There are a number of covert communication techniques such as: Cryptography, Steganography, Covert channel, Anonymity, Watermarking etc. Steganography is one of the effective means of data hiding that protects data from unauthorized or unwanted disclosure. It works by hiding secret messages into ordinary and innocent looking messages those are generally out of suspicion. Digital image Steganography procedures exploit the high capacity and widely used digital images for data hiding purposes [2], [3].

A digital image is a two dimensional function f(x, y) where, x and y are spatial coordinates, f is the amplitude at (x, y), also called the intensity or gray level of the image at that point and x, y, f are finite- discrete quantities. Digital Image processing is the use of computer algorithms to perform image processing on digital images. It allows a wider range of complex and sophisticated algorithms to be applied to digital images with ease and with a much effective way in comparison to analog signal processing [4].

Figure 1, depicts the general block diagram of image Steganography where at the transmitter's end a secret message is embedded to an innocent looking cover image and the resultant stegoimage which is visually same as the original cover is then transmitted over the communication channel without raising any suspicion in the minds of intermediate unintended sniffers/ receivers. At the receiving end the secret message is extracted by the authorized receiver using an extraction algorithm and a valid key. To make this process even more concealed and robust, the message is often encrypted using some encryption technology before embedding and has to be decrypted during extraction to reveal the message.

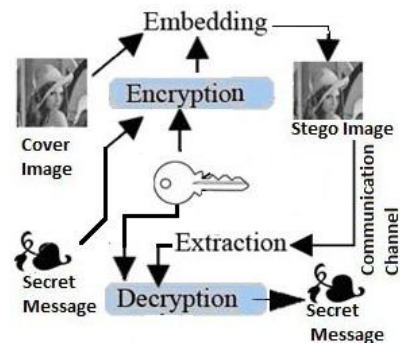

**Figure 1**. Block diagram of Steganography System

There exist both spatial as well as transform domain image Steganography methods. The transform domain procedures are more robust and are commonly used for watermarking purposes whereas the spatial domain methods provide higher capacity and are popular for Steganographic use [5]. LSB substitution is a popular spatial domain method that replaces the lower order image bits those do not carry much useful image information by the secret message bits. In this paper we have used a modified Arnold's Cat Map technique to encrypt the message and the experimental results show that the proposed method provides higher data hiding capacity with improved security and simultaneously preserves the quality of the cover without causing any visual distortion to it.

## 2. ARNOLD'S CAT MAP

Arnold's cat map (ACM) or Arnold transform (AT), proposed by Vladimir Arnold in 1960, is a chaotic map which when applied to a digital image randomizes the original organization of its pixels and the image becomes imperceptible or noisy. However, it has a period p and if iterated p number of times, the original image reappears.

**Definition:** The generalized form of Arnold's cat map can be given by the transformation $\Gamma : T^2 \to T^2$ such that:

$$\begin{pmatrix} x' \\ y' \end{pmatrix} = \begin{pmatrix} 2 & 1 \\ 1 & 1 \end{pmatrix} \begin{pmatrix} x \\ y \end{pmatrix} (\mod N) \quad \ldots\ldots (1)$$

Where, x, y $\in$ {0, 1, 2 … N −1} and N is the size of a digital image [6].

A new image is produced when all the points in an image are manipulated once by equation (1).

Arnold's Cat Map (ACM) is a simple but powerful transform and digital image encryption can be achieved by applying this in the following manner [7]:

Let p be the transform period of an N × N digital image I. Applying ACM for a random iteration of t times (t ∈ [1, p]) to I, a scrambled image Γ is obtained which is completely chaotic and is different from I. Now Γ can be transmitted over the communication channels without revealing any information to the unauthorized receivers or sniffers. At the receiving end the process is repeated for (p − t) times to obtain back the original image. Figure.2 shows the results of Arnold transformation applied to a gray scale Lena image.

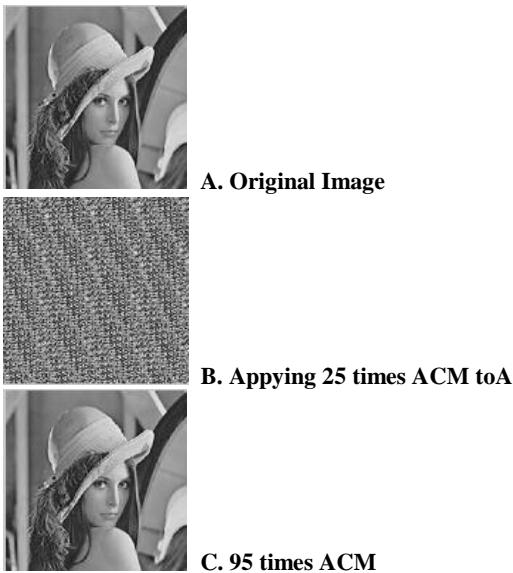

**A. Original Image**

**B. Appying 25 times ACM toA**

**C. 95 times ACM**

**Figure 2.** Arnold's Transformation applied to Lena Image

## 3. MODIFIED ARNOLD TRANSFORM

It can easily be seen that the original Arnold transformations given by equation (1) can be modified to produce a sequence of Arnold transformations as given below:

$$\begin{pmatrix} x' \\ y' \end{pmatrix} = \begin{pmatrix} i & i+1 \\ 1 & 1 \end{pmatrix} \begin{pmatrix} x \\ y \end{pmatrix} (\mod N) \quad \ldots\ldots(2)$$

OR

$$\begin{pmatrix} x' \\ y' \end{pmatrix} = \begin{pmatrix} i+1 & i \\ 1 & 1 \end{pmatrix} \begin{pmatrix} x \\ y \end{pmatrix} (\mod N) \quad \ldots\ldots(3)$$

Where,

$$i \in \{1, 2, 3 \ldots\}$$

Transformations given by equations (2) and (3) are periodic as abs (det (A)) is 1 in both the cases where, A = [a, b; c, d] is the Arnold transform matrix.

Equations (2) and (3), given above, produce a sequence of different Arnold transforms with different periodicity values $P_k$. For example, Fig. 3 shows periodicity of different Arnold Transforms applied to 128x128 grayscale Lena image.

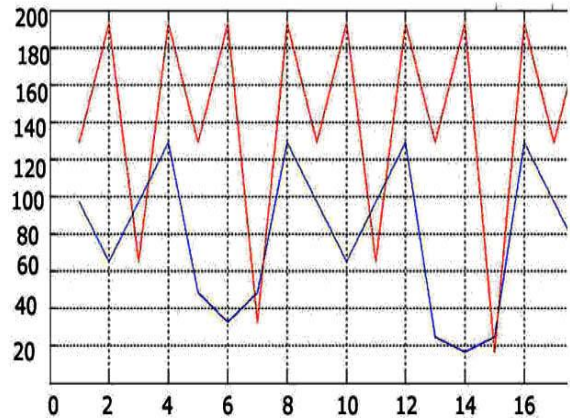

Blue:  x = ((i+1).x + i . y ) mod N , y= (x+y) mod N  i=1…20
Red:   x= (i . x + ( i + 1) . y) mod N, y= (x+y) mod N  i=1…20

**Figure 3.** Periodicity of modified AT for different values of i

From this above picture, it is clear that:

a) the Arnold periodicity varies between 128, 192 and 252 for different i-values for the same 128x128 'micky' image when the first pair of equations are used and when the second pair of equations are used there are 5-different Arnold periodicities between 50 to 252.

b) All these transform functions map the image bits differently.
c) Images scrambled with a particular AT cannot be restored using a deferent AT.

The following example demonstrates the above observations:
Let I be a 3 x 3 matrix given by:

$$I = [1\ 2\ 3;\ 4\ 5\ 6;\ 7\ 8\ 9]$$

Applying the transformation given in equation (1) to I for 1, 2, 3, 4 numbers of times it can be seen that the periodicity in this case is 4.

$$AI1 = [1\ 9\ 5;\ 6\ 2\ 7;\ 8\ 4\ 3]$$
$$AI2 = [1\ 3\ 2;\ 7\ 9\ 8;\ 4\ 6\ 5]$$
$$AI3 = [1\ 5\ 9;\ 8\ 3\ 4;\ 6\ 7\ 2]$$
$$AI4 = [1\ 2\ 3;\ 4\ 5\ 6;\ 7\ 8\ 9]$$

Whereas transforming I through a modified AT given by equation (3) for i=3, we get:

$$BI1 = [1\ 4\ 7;\ 8\ 2\ 5;\ 6\ 9\ 3]$$
$$BI2 = [1\ 8\ 6;\ 9\ 4\ 2;\ 5\ 3\ 7]$$
$$BI3 = [1\ 9\ 5;\ 3\ 8\ 4;\ 2\ 7\ 6]$$
$$BI4 = [1\ 3\ 2;\ 7\ 9\ 8;\ 4\ 6\ 5]$$
$$BI5 = [1\ 7\ 4;\ 6\ 3\ 9;\ 8\ 5\ 2]$$
$$BI6 = [1\ 6\ 8;\ 5\ 7\ 3;\ 9\ 2\ 4]$$
$$BI7 = [1\ 5\ 9;\ 2\ 6\ 7;\ 3\ 4\ 8]$$
$$BI8 = [1\ 2\ 3;\ 4\ 5\ 6;\ 7\ 8\ 9]$$

Now equation (1) when applied to some $BI_i$ above, let say to $BI_2$, it produces:

$$CI1 = [1\ 7\ 4;\ 2\ 8\ 5;\ 3\ 9\ 6]$$
$$CI2 = [1\ 6\ 8;\ 5\ 7\ 3;\ 9\ 2\ 4]$$
$$CI3 = [1\ 4\ 7;\ 3\ 6\ 9;\ 2\ 5\ 8]$$
$$CI4 = [1\ 8\ 6;\ 9\ 4\ 2;\ 5\ 3\ 7]$$

The following properties of AT are clear from the above experiments:

- Both of the transformation functions have different Arnold periodicities (4 in 1st case and 8 in 2nd)
- The scrambling patterns of both are different.
- Applying AT given in equation (1) to any of $BI_i$s, we cannot retrieve back I.

In our proposed Steganography system model, we have exploited the above properties of the Arnold's transformation to make the system more secure against unauthorized access.

## 4. PROPOSED MODEL

The proposed Steganography model has two phases: the embedding phase at the transmitter's end and the extraction phase at the receiver's end. In the embedding phase, the secret message S is first scrambled for some tm (assuming $P_m$ is the period and $0 < t_m < p_m$) number of times using Arnold's cat map at a predefined m different levels, selecting m different transformation functions from equations (2) or equations (3), in a certain order O to make it more secure against unauthorized extraction. This scrambled message S` is embedded into the cover image C to generate the stegoimage C`. C` is transmitted and at the receiving end the secret message(s) is/ are extracted by following the extraction and decryption process in the reverse order. In this technique, the values of i, m, $p_m$, $t_m$, O are kept secret and are only known to the authorized users and extraction without the keys results with noises only, making the procedure secure.

### 4.1 Embedding Algorithm

INPUT: Cover image C of size N x N. Secret messages/Images, let's say; $S_1$, $S_2$, $S_3$ of N x N blocks, Keys: *i, m, $p_m$, $t_m$, O*

For each message/ image $S_i$, do step1 to 3

STEP1: Scramble $S_i$ with some $A_i$ (where Ai is the i[th] AT) for $t_m$ times to obtain Si`
STEP2: Repeated step1 for *m* number of times with different AT say $A_j$ (j !=i) and $t_m$s in order *O* to obtain final scrambled message $S_i^m$.
STEP3: Embed the scrambled messages/ images $S_i^m$s to the LSB planes of the cover image C to get the stegoimage C`.

### 4.2 Extraction Algorithm

INPUT: Stegoimage C` of size N x N.
Keys: *i, m, $p_m$, p`, $t_m$, O*

STEP1: Retrieve $S_i^m$s from the LSBs of C`.
For each $S_i^m$ do:

STEP2: Apply ($p_m$ - $t_m$) times Arnold transforms $A_j$ to obtain Si`
STEP3: Repeat step2 for *m* times with reverse order *O* to get back the secret messages $S_i$s

## 5. EXPERIMENTAL RESULTS AND ANALYSIS

The proposed method is tested and validated over a range of 20 different standard gray scale images of size 128 x128 including 'Lena' and 'Baboon' as the cover images. A number of binary images of size 128 x 128 including 'Logo', 'Micky' and 'Text' are considered as the secret messages/ images for embedding purpose. Figure 4 (B, D) shows the Stego images of original Lena and Baboon image after the secret messages are encrypted using the proposed method and embedded into the LSB, 7th and 6th bit planes of the cover images respectively.

Figure 5 shows the three least significant bit planes of the Lena image, which virtually contains no significant image information and seems like some random noises. Figure 6 shows the information retrieved from the Stego Lena image without using a valid key. It is clear from the figures that the information retrieved without a valid key is completely random, undetectable, unsuspicious and looks like some noise similar to that of figure 5. The secret information retrieved using the valid keys are given in figure 7.

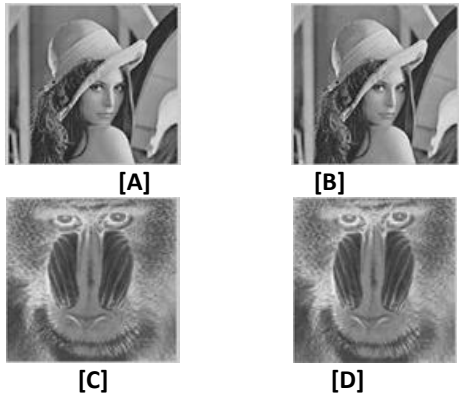

**Figure 4.**
**A, C:** Original Lena & Baboon Images respectively
**B, D:** Lena & Baboon Image after Embedding text into three LSBs

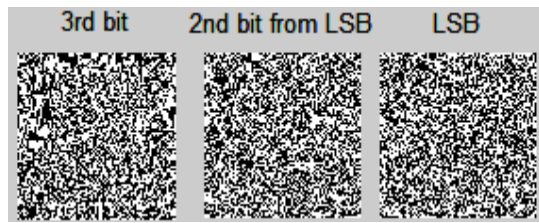

**Figure 5.** Three LSB Bit planes of the Original Lena Image

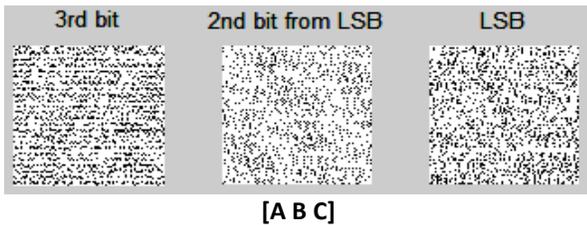

**[A B C]**
**Figure 6.** Retrieval without using valid keys

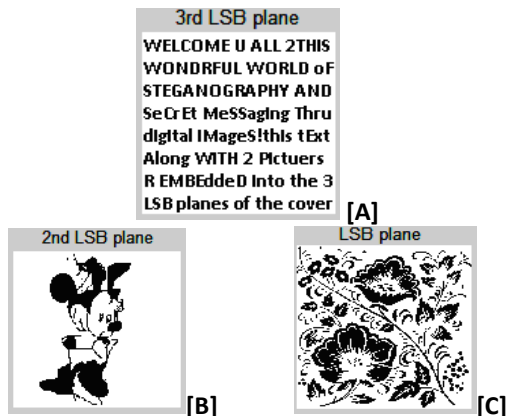

**Figure 7.** Retrieval of information using valid keys [m=2, i=1, 2; pm-tm=100, 162; order: 1, 2]

In case of single Arnold, as Arnold transform is periodic in nature [6], the information can be retrieved by running the algorithm for a certain number of times in random and observing the outputs. For example, figure 2.C, which is same as the original figure 2.A, can be retrieved from figure 2.B even without knowing the period p of the image and p`, the number of times Arnold transform is initially applied to it, just by systematically applying the transform somewhere between 1 to 3p number of times. But in this proposed method the secret information remains highly secured and undetectable as the procedure involves a number of keys. It has been seen that it is not possible to reach at images of figure 7(A, B, C) by applying Arnold transform (AT) to images given in figure 6(A, B, C), a random number of times. Since the original message in this case is nothing but another scrambled image. So, the secret message remains highly secure against hit and trial extractions by unauthorized participants. The data hiding capacity of this method is also much higher in comparison the single LSB substitution method [8], [9]. Table I summarizes the comparison of this method against LSB substitution and simple Arnold Transform methods.

**Table 1. Comparison of Image Steganography Methods**

| Features | Single LSB substitution method | Simple Arnold Transform method | Proposed method |
|---|---|---|---|
| Imperceptibility | Low | Medium | High |
| Capacity | Low | Low | High |
| Robustness | Low | Medium | Medium |
| Encryption | Low | Medium | High |

It has also been observed that the bit preservation ratio of the proposed method is better (most of the times) in comparison to the methods involving unscrambled data insertion. This proves that the distortion to the original image is minimized against the unscrambled three bit substitution methods. The PSNR values after embedding data into 1, 2, 3, 4 bit planes are given in Table II, which shows that the PSNR is as high as 37 dB even with 3-bit insertions.

**Table 2. Text inserted into number of bit planes Vs PSNR**

| Image (128 x 128) size | Embedding data in one Bit Plane | Embedding data in 2 Bit Planes | Embedding data in 3 Bit Planes | Embedding data in 4 Bit Planes |
|---|---|---|---|---|
| Baboon | 51.3797 | 43.6852 | 37.0031 | 30.8176 |
| Lena | 51.0937 | 43.4550 | 36.9229 | 30.4837 |
| Miera | 51.1843 | 43.4743 | 36.9133 | 30.5001 |

## 5. CONCLUSIONS

In spite of availability of a number of Steganography methods, research is still going on to develop methods satisfying all the requirements of Steganography. It is not that an easy task to develop a method that satisfies all the requirements as the requirements may vary with applications. Here we have implemented an algorithm that satisfies both of the attributes such as high imperceptibility & high security. Being a spatial domain method, this of course, is not robust against noise, as the lower order bit planes are generally affected by noises and compression techniques. Future works in this direction include development of some transform domain methods those will provide robustness along with Impeccability, security and insertion into higher order bits to achieve further higher capacity and robustness.